\documentclass{article}
\usepackage{arxiv}
\usepackage[utf8]{inputenc}
\usepackage[T1]{fontenc}
\usepackage{hyperref}
\usepackage{url}
\usepackage{booktabs}
\usepackage{amsfonts}
\usepackage{nicefrac}
\usepackage{microtype}
\usepackage{graphicx}
\usepackage{amsmath}
\usepackage{multirow}
\usepackage{array}
\usepackage{longtable}
\usepackage{tabularx}
\usepackage{float}

\graphicspath{{./}}

\title{The AI Invisibility Effect: Understanding Human-AI Interaction When Users Don't Recognize Artificial Intelligence}

\author{
 Obada Kraishan \\
  College of Media and Communication\\
  Texas Tech University\\
  Lubbock, TX 79409, USA \\
  \url{https://orcid.org/0009-0007-7180-8620} \\
}

\begin{document}
\maketitle

\begin{abstract}
The fast integration of artificial intelligence into mobile applications has completely changed the digital landscape; however, the impact of this change on user perception of AI features remains poorly understood. This large-scale analysis examined 1,484,633 mobile application reviews across 422 applications (200 AI-featuring, 222 control) from iOS App Store and Google Play Store. By employing sentiment classification, topic modeling, and concern-benefit categorization, we identified a major disconnect: only 11.9\% of reviews mentioned AI, even though 47.4\% of applications featured AI capabilities. AI-featuring applications received significantly lower ratings than traditional applications ($d = 0.40$); however, hierarchical regression revealed a hidden pattern---the negative relationship reversed after controlling for AI mentions and review characteristics ($b = 0.405$, $p < .001$). Privacy dominated user concerns (34.8\% of concern-expressing reviews), while efficiency represented the primary benefit (42.3\%). Effects varied greatly by category, from positive for Assistant applications ($d = 0.55$) to negative for Entertainment ($d = -0.23$). These findings suggest that AI features often operate below user awareness thresholds, and it is the explicit recognition of AI, rather than its mere presence, that drives negative evaluations. This challenges basic assumptions about technology acceptance in AI systems.
\end{abstract}

\keywords{artificial intelligence \and human-AI interaction \and mobile applications \and user perception \and technology acceptance \and app reviews \and user experience}

\section{Introduction}

The use of artificial intelligence into everyday digital experiences represents one of the most major technological shifts of the past decade. From voice assistants that interpret natural language to photo editors that enhance images with computational photography, AI has moved from research laboratories into the smartphones of billions of users worldwide \cite{amershi2019,burrell2023}. This transformation has been clear in mobile applications, where AI features have become more common across diverse categories ranging from productivity tools to entertainment platforms \cite{xu2021}. However, despite this widespread deployment, our understanding of how regular users perceive, evaluate, and engage with AI-powered features remains very limited.

The appeal of AI in consumer applications centers on enhanced personalization, improved efficiency, and capabilities that would be impossible through traditional programming approaches \cite{cai2019,yang2020}. Technology companies have invested billions in developing AI features, often positioning them as key differentiators in competitive markets \cite{brynjolfsson2017}. Marketing materials frequently highlight the intelligence and smartness of these applications, suggesting that AI integration automatically improves user experience. However, recent research suggests that the relationship between AI implementation and user satisfaction may be more complex than these narratives suggest \cite{dietvorst2015,longoni2019}.

Multiple areas of research have examined different aspects of human-AI interaction, though significant gaps remain in our understanding of real-world user responses to AI features in consumer applications. Studies in human-computer interaction have primarily focused on controlled laboratory experiments or prototype systems \cite{amershi2019,kocielnik2019}, while research in information systems has examined organizational AI adoption rather than individual consumer experiences \cite{dwivedi2021,rai2019}. Marketing research has investigated consumer attitudes toward AI in specific contexts such as customer service or recommendation systems \cite{huang2021,mende2019}. However, these approaches typically examine attitudes in controlled settings rather than capturing naturalistic user feedback from real-world application usage.

This work addresses these limitations by analyzing a large collection of real user feedback from app store reviews, providing insights into how millions of users experience and evaluate AI features in their daily digital use. App store reviews offer a unique window into authentic user sentiment, as they represent voluntary, unprompted feedback generated in real-world usage contexts \cite{harman2012,pagano2013}. Unlike laboratory studies or surveys, these reviews capture user reactions to fully deployed commercial applications, reflecting the complexities and constraints of actual implementation rather than simplified test versions.

The timing of this research is critical due to the recent acceleration in AI deployment, which follows breakthroughs in large language models and generative AI systems \cite{brown2020,bubeck2023}. The period from 2024 to 2025 is expected to witness huge public attention to AI capabilities, with tools like ChatGPT and Claude becoming household names \cite{haque2022}. As public attention to AI capabilities grows, important questions arise about whether and how users recognize AI features in their everyday applications. Building on these observations, this study addresses three research questions that collectively illuminate the current state of AI integration in consumer mobile applications:

\begin{itemize}
\item \textbf{RQ1:} How do AI-featuring applications differ from non-AI applications in terms of user satisfaction and engagement patterns?
\item \textbf{RQ2:} What specific concerns and perceived benefits do users express when discussing AI features in mobile applications?
\item \textbf{RQ3:} To what extent does the reception of AI features vary across different application categories and mobile platforms?
\end{itemize}

This research makes several important contributions to our understanding of human-AI interaction in consumer contexts. First, we provide a large-scale analysis of user responses to AI-powered mobile applications in real-world settings, examining over 1.48 million unprompted reviews across 422 applications from both iOS and Android platforms. Second, we examine whether and how users recognize and articulate AI features in their naturalistic evaluations. Third, we identify complex relationships between AI implementation and user satisfaction through hierarchical regression analyses. Fourth, we develop a comprehensive taxonomy of user concerns and perceived benefits related to AI features. Finally, we demonstrate substantial heterogeneity in AI reception across application categories and platforms.

\section{Literature Review}

\subsection{The Evolution of AI in Consumer Applications}

The use of artificial intelligence into consumer-facing applications has developed through several different phases, each characterized by different technological capabilities and user expectations. Early implementations focused on rule-based systems and simple pattern recognition, offering features like predictive text and basic recommendation algorithms \cite{kaplan2019}. These initial attempts into AI were often invisible to users, operating as background processes that enhanced functionality without clear user awareness. The introduction of virtual assistants like Siri in 2011 marked a big shift toward more visible and interactive AI features, though adoption remained limited by technical constraints and user doubts \cite{cowan2017}.

The current generation of AI-powered applications, appearing after 2020, represents a qualitative advance in both capability and widespread use. Advances in deep learning, mainly in natural language processing and computer vision, have enabled applications to perform tasks previously thought to require human intelligence \cite{lecun2015,vaswani2017}. Photo editing applications can now automatically enhance images using computational photography techniques that surpass manual editing for most users \cite{gharbi2017}. Writing assistants powered by large language models can generate, edit, and translate text with near-human fluency \cite{brown2020}. These capabilities have led to a transformation of mobile applications where AI features have become expected rather than exceptional \cite{dwivedi2021}.

Despite these technical advances, research on user reception of AI features reveals a complex and often inconsistent landscape. Studies have documented both enthusiasm for AI capabilities and deep skepticism about their implementation \cite{cave2019,glikson2020}. Research shows that users often prefer human judgment even when algorithmic decisions clearly perform better than human alternatives \cite{dietvorst2015}. On the other hand, other research identifies situations where users over-rely on AI recommendations, potentially leading to loss of skills or reduced critical thinking \cite{parasuraman2010}. These conflicting findings suggest that user responses to AI features depend heavily on context, implementation, and individual differences. However, while prior research has examined AI acceptance through laboratory experiments, surveys, or studies of specific AI systems in isolation, large-scale empirical analysis of how users evaluate AI features across real App Store and Google Play reviews from multiple categories and platforms remains limited. This gap is important given that App Store and Google Play reviews represent authentic, voluntary user feedback about fully deployed commercial applications rather than prototype systems or hypothetical scenarios.

\subsection{User Perceptions and the AI Paradox}

A growing body of literature has identified a paradox in consumer responses to AI---the simultaneous desire for AI features and concern over their implications. Users show enthusiasm for the efficiency and convenience that AI promises while at the same time expressing concerns over privacy, personal autonomy, and the authenticity of AI-mediated experiences \cite{sundar2020}. This paradox manifests differently across demographics, with younger users characteristically more accepting of AI features compared to older users who tend to be more skeptical \cite{gursoy2019}. 

Importantly, existing studies of AI acceptability typically assume that users consciously recognize and can differentiate the AI features they are evaluating. However, research examining whether users actually recognize AI in consumer applications, or whether AI features might shape user experience without explicit awareness, remains limited. This assumption of conscious awareness deserves scrutiny, as AI increasingly operates invisibly within applications, embedded in features users may not recognize as artificially intelligent.

Trust emerges as a central factor mediating user acceptance of AI features. Research in human-computer interaction has identified multiple dimensions of trust in AI systems, including competence trust, benevolence trust, and integrity trust \cite{hoff2015,jacovi2021}. The development of trust in AI systems follows different patterns than trust in human agents or traditional software, with users often struggling to develop appropriate mental models of AI capabilities and limitations \cite{bansal2019}.

The transparency of AI systems presents another critical factor influencing user perceptions. While calls for explainable AI have become prominent in both research and policy discussions \cite{arrieta2020,gunning2019}, studies suggest that users may not always desire or benefit from detailed explanations of AI operations. Research by Kizilcec \cite{kizilcec2016} identified a transparency paradox where too much information about algorithmic processes can reduce trust and satisfaction. This finding complicates simplistic assumptions about the relationship between AI transparency and user acceptance, suggesting that effective AI integration requires careful consideration of what information users want and can meaningfully process.

\subsection{Privacy Concerns and Data Anxieties}

Privacy concerns represent perhaps the most consistently documented user apprehension regarding AI features in consumer applications. The data-intensive nature of machine learning systems requires extensive collection and processing of user information, raising questions about data security, usage, and control \cite{acquisti2015,martin2017}. Studies have shown that users often lack awareness of the extent of data collection required for AI features, leading to what Barocas and Nissenbaum \cite{barocas2014} term context collapse: the breakdown of expected boundaries between different spheres of personal information.

The privacy calculus framework suggests that users engage in a cost-benefit analysis when deciding whether to use AI-powered features, weighing perceived benefits against privacy risks \cite{dinev2006,smith2011}. However, empirical research indicates that this calculus is often flawed, with users demonstrating inconsistent preferences and behaviors: the so-called privacy paradox where stated concerns do not align with actual usage patterns \cite{kokolakis2017}. In the context of AI applications, this paradox may be exacerbated by the opacity of data processing and the difficulty users face in understanding how their information is used to train and improve AI systems.

Recent research has also highlighted concerns about inferential privacy: the ability of AI systems to deduce sensitive information from seemingly innocuous data \cite{wachter2019}. Users may be comfortable sharing certain types of information but uncomfortable with the inferences AI systems can draw from that data. This concern is particularly acute in applications that use AI for personalization or prediction, where the system's ability to anticipate user needs may be perceived as intrusive rather than helpful \cite{tene2013}.

\subsection{Platform and Category Differences}

The mobile application ecosystem is characterized by significant heterogeneity across platforms and application categories, factors that may substantially influence user reception of AI features. Research comparing iOS and Android users has identified systematic differences in user demographics, technology adoption patterns, and privacy preferences \cite{bohmer2011,petsas2013}. iOS users tend to have higher income levels and show greater willingness to pay for applications, while Android users represent a more diverse global population with varying technical literacy levels \cite{statista2023}. These platform differences may translate into different expectations and evaluations of AI features.

Application category represents another important source of variation in AI reception. Research in information systems has long recognized that technology acceptance varies by task characteristics and usage context \cite{venkatesh2003}. In the context of AI features, users may have different expectations and tolerance levels depending on whether they are using productivity tools, entertainment applications, or creative platforms. For instance, AI assistance in productivity applications may be evaluated primarily on efficiency grounds, while AI in creative applications may raise concerns about authenticity and human agency \cite{chamberlain2023,epstein2023}.

The concept of AI readiness at the category level suggests that some application domains may be more suitable for AI integration than others. Research by Jussupow et al. \cite{jussupow2021} identifies factors such as task complexity, outcome measurability, and error tolerance as determinants of successful AI implementation. Entertainment applications, for example, may have lower error tolerance for AI features that disrupt user enjoyment, while utility applications may benefit from AI automation of routine tasks. Understanding these category-specific dynamics is essential for explaining variation in user responses to AI features.

\subsection{Theoretical Framework}

Drawing on these diverse literature streams, we adopt a multi-theoretical perspective that combines insights from technology acceptance models, trust theory, and privacy calculus frameworks. The Technology Acceptance Model (TAM) and its extensions provide a foundation for understanding how perceived usefulness and ease of use influence user evaluation of AI features \cite{davis1989,venkatesh2000}. However, traditional TAM constructs may be insufficient for capturing the unique characteristics of AI systems, particularly concerns about autonomy, authenticity, and algorithmic decision-making.

We therefore augment TAM with concepts from trust theory, recognizing that user acceptance of AI features depends not only on functional evaluations but also on trust in the system's competence, benevolence, and integrity \cite{mcknight2002}. The automation trust framework developed by Lee and See \cite{lee2004} provides additional insights into how users calibrate their reliance on AI systems, identifying factors that lead to both over-trust and under-trust. This framework is particularly relevant for understanding the disconnect between AI capabilities and user utilization patterns observed in many consumer applications.

Privacy calculus theory offers a lens for understanding how users navigate the trade-offs inherent in AI-powered features \cite{culnan1999,xu2011}. We extend traditional privacy calculus models to account for the unique characteristics of AI systems, including the opacity of data processing, the potential for inferential privacy violations, and the difficulty users face in assessing privacy risks. This extended framework helps explain why users may simultaneously express privacy concerns while continuing to use AI-powered features, and why privacy salience varies across different application contexts.

\section{Method}

\subsection{Data Collection and Sampling}

This study analyzed publicly available user reviews from mobile application marketplaces using a large-scale observational design. Data collection occurred over a more extended period to ensure comprehensive data coverage, utilizing automated web scraping with appropriate rate limiting to comply with platform guidelines. Applications were systematically selected based on download numbers and user ratings from the top 500 free applications on both Apple App Store and Google Play Store. Three inclusion criteria were applied: (1) sustained presence in the top charts over three months, indicating consistent user engagement; (2) availability on both iOS and Android platforms to enable cross-platform comparisons; and (3) a minimum of 1,000 user reviews to ensure adequate statistical power for application-level analyses.

AI-featuring applications were identified through a weighted analysis of multiple indicators: explicit AI-related terminology in descriptions, analysis of release notes documenting AI feature introduction, and verification through technology news sources and developer announcements. This multi-source approach identified 200 applications with confirmed AI features, including specific examples such as recommendation algorithms, natural language processing, and generative AI capabilities. The remaining 222 control applications showed no evidence of AI features and were matched to AI applications using a detailed methodology based on category, user base size, and release timeframe to minimize confounding variables.

\subsection{Sample Characteristics}

The final dataset, collected between January 2024 and January 2025, comprised 1,484,633 reviews from 422 applications (200 AI-featuring, 222 control). Reviews were unevenly distributed across platforms (Android: 72.2\%, n = 1,072,246; iOS: 27.8\%, n = 412,387), which may reflect marketplace dynamics. Applications spanned 23 categories, with the strongest representation in Productivity (18.2\%), Entertainment (14.5\%), and Creative (12.8\%), indicating a focus on these areas. The average application generated 3,518 reviews, with a standard deviation of 4,892, indicating significant variability in review counts. Table~\ref{tab:sample} presents detailed sample characteristics, including distribution across categories and platforms.

\begin{table}[h]
\centering
\caption{Sample Characteristics and Distribution}
\begin{tabularx}{\columnwidth}{Xl}
\toprule
\textbf{Characteristic} & \textbf{n (\%)} \\
\midrule
Total reviews & 1,484,633 \\
Total applications & 422 \\
\quad AI-featuring apps & 200 (47.4\%) \\
\quad Control apps & 222 (52.6\%) \\
Reviews with AI mentions & 176,658 \\
Percentage of total & 11.9\% \\
\textbf{Platform} & \\
\quad iOS & 412,387 (27.8\%) \\
\quad Android & 1,072,246 (72.2\%) \\
Time period & Jan 2024 - Jan 2025 \\
Average review length (words) & 47.3 ($SD$ = 68.2) \\
Average rating (overall) & 3.58 ($SD$ = 1.52) \\
\bottomrule
\end{tabularx}
\label{tab:sample}
\end{table}

\subsection{Data Extraction Procedures}

Data extraction was conducted in three distinct phases, each utilizing specialized Python libraries, which will be detailed in the following sections. Application metadata was collected using App Store Scraper (v1.0.0) for iOS and Google Play Scraper (v1.2.0) for Android, extracting standardized fields including identifiers, developer information, categories, version histories, ratings, and descriptions.

Review collection used tailored strategies for each platform to effectively address API limitations, such as combining direct scraping with iTunes RSS feeds for iOS. For iOS reviews, a multi-method approach was adopted that combined direct scraping, iTunes RSS feeds, and the iTunes Lookup API as fallbacks. Android reviews were collected through the Google Play Scraper API with pagination support. Collection parameters, such as a maximum of 1,000 reviews per application and a minimum length of 10 characters, were chosen to ensure a manageable dataset size while maintaining data quality. Each review record captured text content, rating (1--5 stars), submission date, application version, and developer responses when available.

Release notes and version histories were analyzed to pinpoint the timing of AI feature introductions, providing insights into the development timeline. The iTunes Lookup API provided iOS release notes, while Android data came from the recentChanges field. In cases where release notes were unavailable, application descriptions were examined to identify AI features, serving as an alternative method for feature identification.

\subsection{Privacy and Ethical Considerations}

All data collection adhered to ethical standards for publicly available content. Reviewer names were replaced with MD5-hashed identifiers to prevent re-identification while preserving analytical uniqueness. Review texts were automatically scanned and redacted for personal information, including email addresses, phone numbers, and real names. Platform terms of service were respected through rate limiting, and no attempts were made to access restricted content or circumvent platform restrictions.

\subsection{Measures and Operationalization}

AI-related content detection utilized a dictionary of 74 terms and phrases, which were derived from academic literature and refined through manual review, to identify relevant content. The dictionary spanned three categories, each relevant to the detection process: core AI concepts (e.g., ``artificial intelligence,'' ``machine learning''), specific models (e.g., ``GPT,'' ``Claude''), and feature implementations (e.g., ``AI assistant,'' ``smart reply''). Each review received binary classification for AI mention presence and a continuous measure of keyword density per 100 words, which contributed to assessing the prevalence and intensity of AI-related content.

Sentiment analysis combined rule-based and machine learning approaches. The rule-based component used weighted dictionaries of positive and negative terms. The machine learning component implemented a DistilBERT model that was fine-tuned on product review data, providing probability distributions across sentiment categories. The transformer model served as the primary classifier when its confidence exceeded 0.7, while the rule-based system was used to validate the results of the transformer model.

Content analysis identified concerns and benefits through pattern matching against validated dictionaries developed from qualitative review analysis. Eight concern and eight benefit categories were operationalized, with phrase matches receiving double weight compared to individual keywords to reflect their greater specificity and relevance in context.

\subsection{Statistical Analysis}

We conducted statistical analyses using Python, primarily relying on the pandas library for data manipulation, scipy for statistical tests, statsmodels for regression analyses, and scikit-learn for machine learning components. The analysis plan was pre-registered to minimize researcher degrees of freedom, with exploratory analyses conducted post-hoc clearly identified in the results.

Following the exploratory analyses, group comparisons between AI-featuring and non-AI applications were conducted using independent samples t-tests with Welch's correction to account for unequal variances between groups. Chi-square tests examined categorical comparisons, including rating distributions and AI mention patterns. Effect sizes were calculated using Cohen's d for continuous outcomes. For categorical associations, we computed Cramér's V as an effect size measure that accounts for table dimensions.

Multiple regression models examined the relationship between AI features and user ratings, while controlling for potential confounding variables. The base model included AI application status as the primary predictor with user rating as the outcome. Subsequent models added controls for application category, platform, review length, and temporal factors. We tested for interaction effects between AI status and other predictors, particularly platform and category, to identify potential moderating factors.

Logistic regression models examined factors predicting the probability of five-star reviews, with odds ratios calculated to quantify effect magnitudes. Analysis of variance (ANOVA) tested for differences across multiple application categories, with eta-squared ($\eta^2$) reported as the effect size measure. Two-way ANOVA examined interaction effects between platform and AI status. Post-hoc pairwise comparisons with Bonferroni correction identified specific category differences. Pearson correlation coefficients assessed associations between specific concerns/benefits and ratings. For temporal trend analysis, the Cochran-Armitage test evaluated monotonic increases in privacy concerns over the study period. All models reported appropriate fit statistics, including $R^2$ for linear models and pseudo-$R^2$ for logistic models.

In addition to statistical models, topic modeling employed Latent Dirichlet Allocation (LDA) to identify thematic patterns within AI-mentioning reviews. Text preprocessing involved tokenization using the spaCy library, lemmatization to reduce words to their base forms, removal of stop words using an extended list that included domain-specific terms, and extraction of both unigrams and bigrams to capture multi-word concepts. To support the topic modeling process, the document-term matrix was constructed using term frequency-inverse document frequency (TF-IDF) weighting with a maximum vocabulary of 1,000 features to balance comprehensiveness with computational efficiency. Model selection employed a grid search over $k$ = 5 to 20 topics, evaluating models based on perplexity scores and topic coherence metrics. The final model with $k$ = 10 topics provided an optimal balance between interpretability and statistical fit.

\section{Results}

The following sections present detailed findings addressing each research question, examining satisfaction differences, user concerns and benefits, and heterogeneous effects across platforms and categories.

\subsection{RQ1: AI-Featuring versus Non-AI Applications}

AI-featuring applications received substantially lower average ratings ($M$ = 3.28, $SD$ = 1.54) than control applications ($M$ = 3.87, $SD$ = 1.43), $t$(420) = $-$4.13, $p$ < .001, $d$ = 0.40. This moderate effect persisted across analytical approaches. AI applications showed lower proportions of five-star reviews (32.4\% vs. 41.8\%) and higher proportions of one-star reviews (24.7\% vs. 16.2\%), $\chi^2$(4, $N$ = 1,484,633) = 18,942.31, $p$ < .001, Cramér's $V$ = .113. Review length also differed, with AI applications generating longer feedback ($M$ = 52.1 words) than non-AI applications ($M$ = 42.5 words), $t$(1,484,631) = 47.93, $p$ < .001, $d$ = 0.14. Table~\ref{tab:comparative} summarizes comparative statistics.

\begin{table}[h]
\centering
\caption{Comparative Statistics for AI-Featuring Versus Control Applications}
\begin{tabularx}{\columnwidth}{Xll}
\toprule
\textbf{Measure} & \textbf{Value} & \textbf{95\% CI} \\
\midrule
\textbf{Rating, M (SD)} & & \\
\quad AI-featuring apps & 3.28 (1.54) & [3.26, 3.30] \\
\quad Control apps & 3.87 (1.43) & [3.85, 3.89] \\
\quad Difference & -0.59*** & [-0.62, -0.56] \\
\quad Cohen's d & 0.40 & [0.37, 0.43] \\
\textbf{5-star reviews, \%} & & \\
\quad AI-featuring apps & 32.4 & [32.1, 32.7] \\
\quad Control apps & 41.8 & [41.5, 42.1] \\
\textbf{1-star reviews, \%} & & \\
\quad AI-featuring apps & 24.7 & [24.4, 25.0] \\
\quad Control apps & 16.2 & [15.9, 16.5] \\
\textbf{Review length (words), $M$ ($SD$)} & & \\
\quad AI-featuring apps & 52.1 (71.3) & [51.7, 52.5] \\
\quad Control apps & 42.5 (64.8) & [42.1, 42.9] \\
\bottomrule
\end{tabularx}
\label{tab:comparative}
\end{table}

\begin{figure}[h]
\centering
\includegraphics[width=0.8\columnwidth]{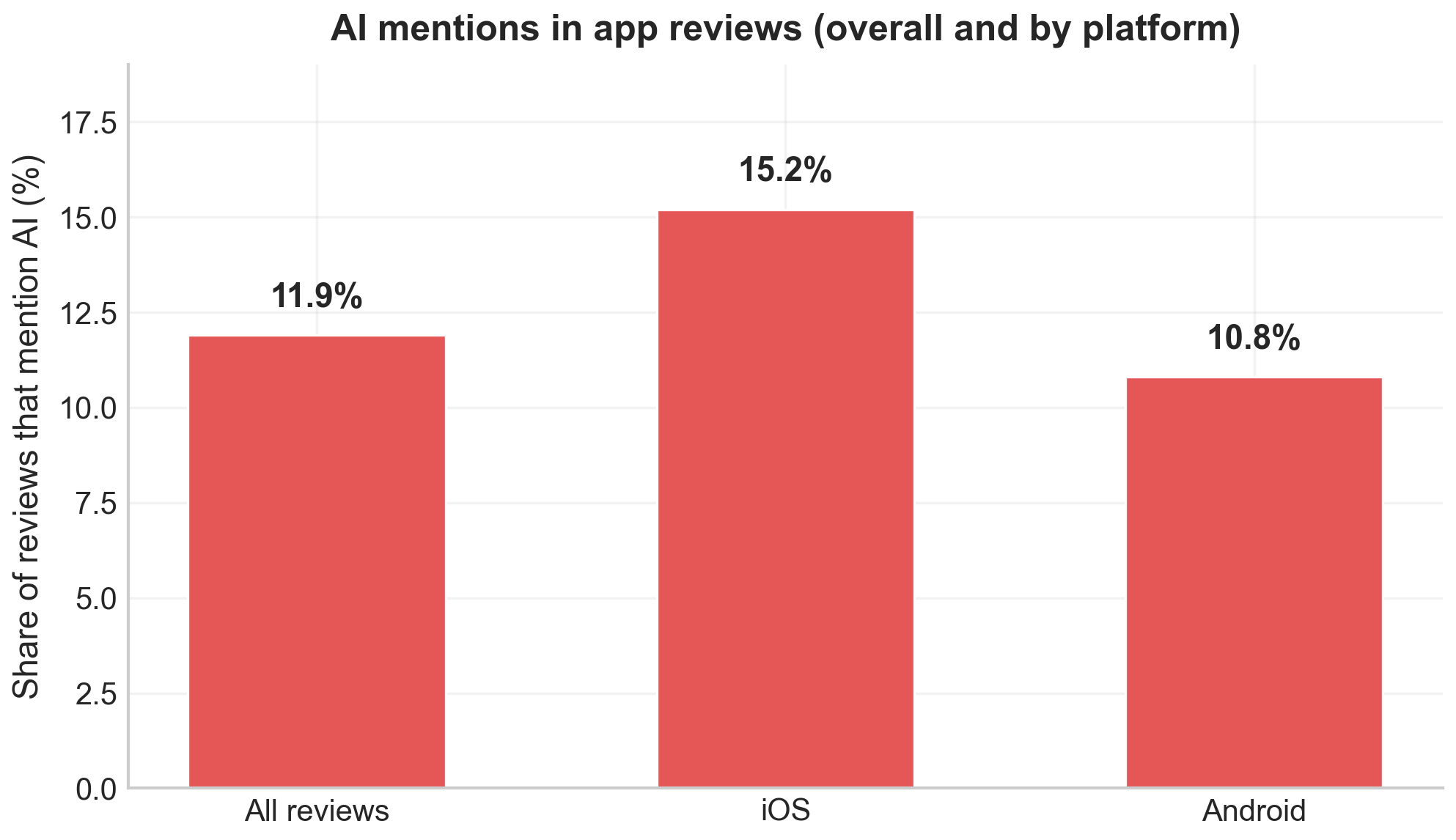}
\caption{AI Mentions in App Reviews}
\label{fig:ai_mentions}
\end{figure}

Although nearly half of apps used AI, only 11.9\% of reviews mentioned it---24\% within AI apps versus 0.8\% in controls ($\phi$ = .34, $p$ < .001). Mentions were somewhat higher on iOS than Android (15.2\% vs. 10.8\%; $\chi^2$(1, $N$ = 1,484,633) = 6,234.45, $p$ < .001; see Figure~\ref{fig:ai_mentions} and Table~\ref{tab:ai_mentions}).

\begin{table}[h]
\centering
\caption{AI Mention Patterns and Distribution}
\begin{tabularx}{\columnwidth}{Xll}
\toprule
\textbf{Pattern} & \textbf{Frequency} & \textbf{Effect Size} \\
\midrule
\textbf{Reviews mentioning AI (\%)} & 176,658 (11.9\%) & \\
\quad In AI apps & 23.7\% & $\phi$ = 0.34*** \\
\quad In non-AI apps & 0.8\% & \\
\textbf{Sentiment in AI mentions} & & \\
\quad Positive only & 45.2\% & \\
\quad Negative only & 31.8\% & \\
\quad Mixed & 16.3\% & \\
\quad Neutral & 6.7\% & \\
\textbf{Rating by mention type (M $\pm$ SD)} & & \\
\quad With AI mention & 3.42 $\pm$ 1.61 & \\
\quad Without AI mention & 3.59 $\pm$ 1.49 & \\
\quad Difference & -0.17*** & d = 0.11 \\
\textbf{Top AI keywords (\%)} & & \\
\quad ``AI'' & 68.4\% & \\
\quad ``ChatGPT'' & 24.7\% & \\
\quad ``bot'' & 18.3\% & \\
\quad ``smart'' & 15.2\% & \\
\quad ``intelligent'' & 12.1\% & \\
\bottomrule
\end{tabularx}
\label{tab:ai_mentions}
\end{table}

\begin{figure*}[t]
\centering
\includegraphics[width=0.9\textwidth]{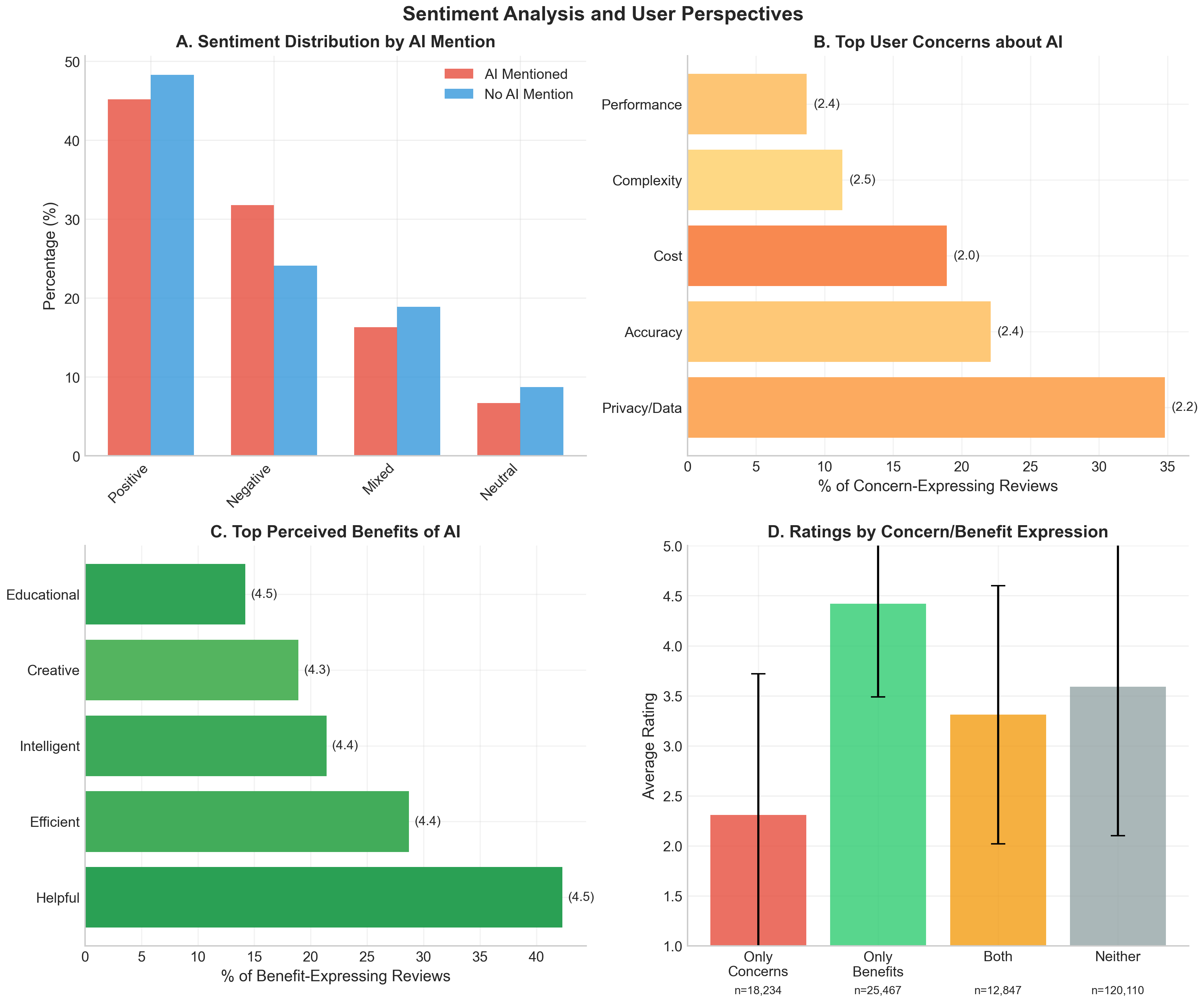}
\caption{Sentiment Analysis and User Perspectives}
\label{fig:sentiment}
\end{figure*}

Hierarchical regression analysis examined factors predicting user ratings (Table~\ref{tab:regression}). Model 1 showed AI applications received lower ratings ($b$ = $-$0.59, $p$ < .001). Adding AI mentions (Model 2) revealed both factors independently predicted lower ratings. The minimal interaction effect (Model 3: $b$ = 0.03, $p$ = .464) suggests independent rather than synergistic effects. Critically, in the full model (Model 4), the AI application coefficient reversed sign ($b$ = 0.405, $p$ < .001) after including controls, with the model explaining $R^2$ = .084 of variance, $F$(23, 99,971) = 397.83, $p$ < .001. This pattern indicates a suppression effect whereby negative bivariate associations are masked by AI apps attracting longer, more critical reviews.

\begin{table}[h]
\centering
\caption{Hierarchical Regression Models Predicting Review Ratings}
\small
\begin{tabularx}{\columnwidth}{Xcccc}
\toprule
\textbf{Variable} & \textbf{M1} & \textbf{M2} & \textbf{M3} & \textbf{M4} \\
\midrule
Intercept & 3.29*** & 3.29*** & 3.29*** & 3.13*** \\
 & (0.01) & (0.01) & (0.01) & (0.01) \\
AI App & $-$0.59*** & $-$0.48*** & $-$0.48*** & 0.41*** \\
 & (0.02) & (0.02) & (0.02) & (0.03) \\
AI Mention & & $-$0.18*** & $-$0.18*** & $-$0.22*** \\
 & & (0.02) & (0.02) & (0.02) \\
AI $\times$ Mention & & & 0.03 & \\
 & & & (0.04) & \\
Platform & & & & 0.00 \\
 & & & & (0.01) \\
Log(Reviews) & & & & 0.06*** \\
 & & & & (0.00) \\
Word Count & & & & $-$0.02*** \\
 & & & & (0.00) \\
$R^2$ & .012 & .012 & .012 & .084 \\
Adj. $R^2$ & .012 & .012 & .012 & .084 \\
$N$ & 100,000 & 100,000 & 100,000 & 100,000 \\
\bottomrule
\end{tabularx}
\smallskip

\raggedright
\small\textit{Note.} M1 = Base; M2 = + AI Mention; M3 = Interaction; M4 = Full Model. Unstandardized $b$ (SE). DV = star rating (1--5). ***$p$ < .001.
\label{tab:regression}
\end{table}

Logistic regression predicting five-star reviews corroborated these findings (Table~\ref{tab:logistic}). AI applications initially showed lower odds of five-star ratings ($OR$ = 0.67, 95\% CI [0.65, 0.69], $p$ < .001) but demonstrated higher odds after adding controls ($OR$ = 1.55, 95\% CI [1.51, 1.59], $p$ < .001). Reviews mentioning AI remained less likely to receive five stars ($OR$ = 0.75, 95\% CI [0.73, 0.77], $p$ < .001).

\begin{table}[h]
\centering
\caption{Logistic Regression Model Predicting 5-Star Reviews}
\begin{tabularx}{\columnwidth}{Xllll}
\toprule
\textbf{Variable} & \textbf{B} & \textbf{SE} & \textbf{OR} & \textbf{95\% CI} \\
\midrule
Intercept & -0.90*** & 0.01 & 0.41 & [0.41, 0.41] \\
AI App (0/1) & +0.44*** & 0.02 & 1.55 & [1.51, 1.59] \\
AI Mention (0/1) & -0.29*** & 0.02 & 0.75 & [0.73, 0.77] \\
Platform (iOS) & +0.08* & 0.04 & 1.08 & [1.00, 1.16] \\
Log(Total Reviews) & +0.06*** & 0.00 & 1.06 & [1.06, 1.06] \\
Pseudo R2 & 0.007 & & & \\
N & 100,000 & & & \\
\bottomrule
\end{tabularx}
\label{tab:logistic}
\end{table}

These findings demonstrate that AI-featuring applications differ significantly from non-AI applications in satisfaction and engagement patterns. The striking gap between AI presence (47.4\% of apps) and explicit user awareness (11.9\% of reviews) suggests users rarely recognize or articulate AI features. The suppression effect indicates that apparent negative reception stems from reviewers who explicitly mention AI rather than AI presence per se. When controlling for AI mentions and review characteristics, AI applications show higher odds of five-star ratings, suggesting implementation quality and user awareness---rather than AI technology itself---drive satisfaction differences.

\subsection{RQ2: User Concerns and Perceived Benefits}

Users report both concerns and benefits around AI features. AI-mentioning reviews are more polarized and slightly lower-rated than non-AI reviews. Among 176,658 AI-mentioning reviews, 32.1\% expressed concerns, 45.3\% benefits, and 29.0\% both. Sentiment was more polarized than in non-AI reviews (45.2\% strongly positive; 31.8\% strongly negative vs. 33.6\% and 24.1\%), $\chi^2$(2, $N$ = 50,000) = 157.59, $p$ < .001, Cramér's $V$ = .056. Reviews that mentioned AI were rated slightly lower ($M$ = 3.42, $SD$ = 1.61) than those that did not ($M$ = 3.59, $SD$ = 1.49), $t$(1,484,631) = $-$43.52, $p$ < .001, $d$ = 0.11, and showed higher capitalization ratios. Figure~\ref{fig:sentiment} presents a four-panel analysis including (A) sentiment distribution comparing AI-mentioned versus non-AI reviews; (B) top user concerns about AI features; (C) top perceived benefits of AI features; and (D) average ratings by concern/benefit expression patterns with error bars showing standard deviations.

Top concerns were privacy/data (34.8\%), accuracy/errors (22.1\%), and cost/subscription (18.9\%); top benefits were helpfulness/utility (42.3\%) and efficiency/time-saving (28.7\%) (Table~\ref{tab:concerns_benefits}).

\begin{table}[h]
\centering
\caption{User Concerns and Benefits in AI-Mentioning Reviews}
\begin{tabularx}{\columnwidth}{Xcc}
\toprule
\textbf{Category} & \textbf{\% of AI} & \textbf{Avg} \\
 & \textbf{Reviews} & \textbf{Rating} \\
\midrule
\textbf{CONCERNS (share among} & & \\
\textbf{concern-expressing reviews)} & 32.1 & 2.31 \\
\hspace{1em} Privacy/Data & 34.8 & 2.18 \\
\hspace{1em} Accuracy/Errors & 22.1 & 2.42 \\
\hspace{1em} Cost/Subscription & 18.9 & 1.97 \\
\hspace{1em} Complexity/Usability & 11.3 & 2.54 \\
\hspace{1em} Performance/Technical & 8.7 & 2.38 \\
\hspace{1em} Ethics/Bias & 3.1 & 2.89 \\
\hspace{1em} Dependency/Overreliance & 0.8 & 3.12 \\
\hspace{1em} Authenticity/Fakeness & 0.3 & 2.76 \\
\textbf{BENEFITS (share among} & & \\
\textbf{benefit-expressing reviews)} & 45.3 & 4.42 \\
\hspace{1em} Helpful/Useful & 42.3 & 4.51 \\
\hspace{1em} Efficient/Time-saving & 28.7 & 4.38 \\
\hspace{1em} Intelligent/Smart & 21.4 & 4.42 \\
\hspace{1em} Creative/Innovative & 18.9 & 4.29 \\
\hspace{1em} Educational/Learning & 14.2 & 4.48 \\
\hspace{1em} Accessible/Easy & 13.6 & 4.31 \\
\hspace{1em} Personalized/Adaptive & 7.8 & 4.27 \\
\hspace{1em} Entertaining/Fun & 6.9 & 4.19 \\
\bottomrule
\end{tabularx}
\label{tab:concerns_benefits}
\end{table}

\begin{figure*}[t]
\centering
\includegraphics[width=0.9\textwidth]{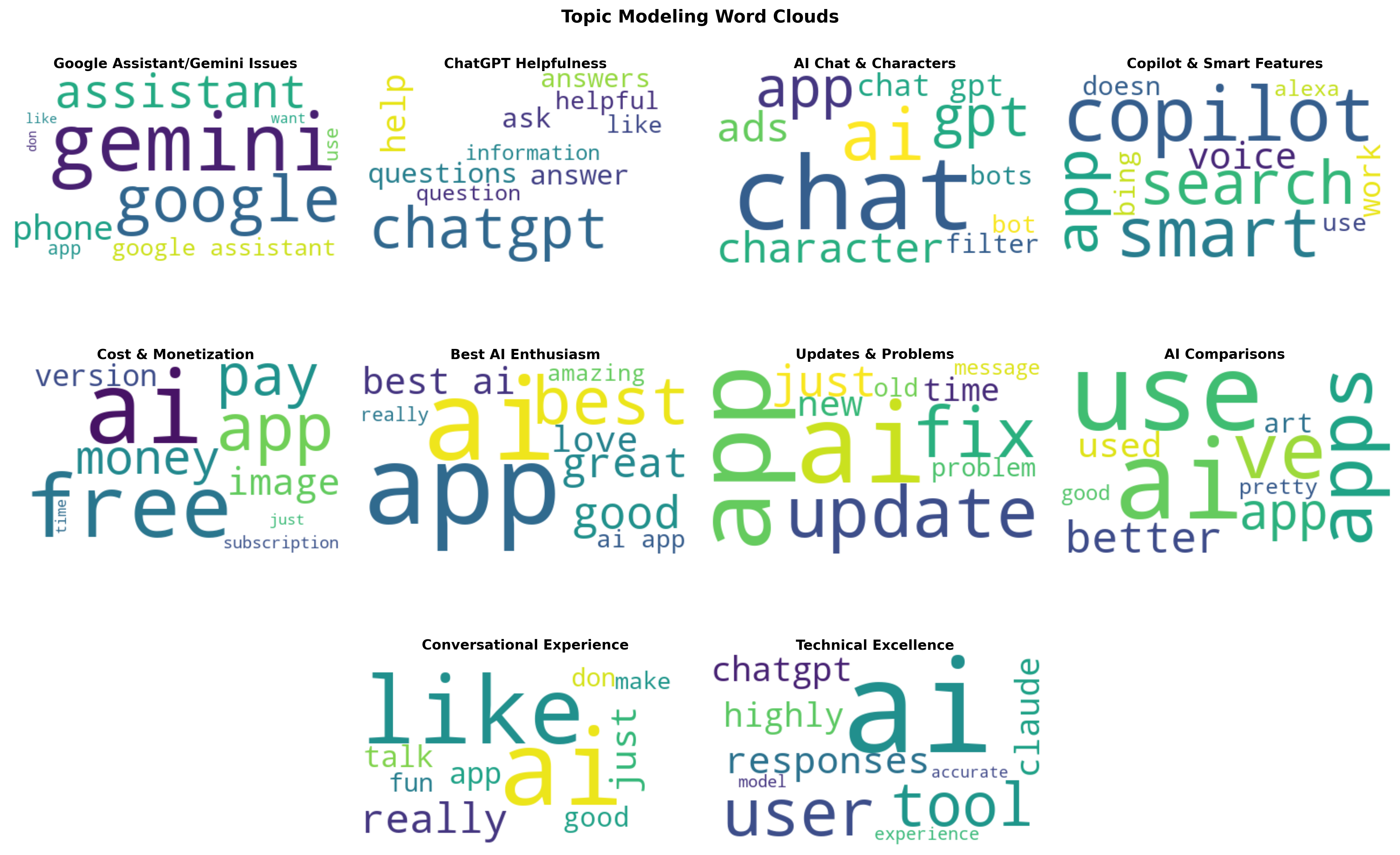}
\caption{Topic Modeling Word Clouds}
\label{fig:topics}
\end{figure*}

Independent-samples t-tests showed concern-only reviews were much lower-rated than benefit-only reviews ($\Delta$ = 2.11; $M$ = 2.31, $SD$ = 1.41 vs. $M$ = 4.42, $SD$ = 0.93), $t$(43,701) = 89.23, $p$ < .001, $d$ = 1.77. Mixed reviews were intermediate ($M$ = 3.31, $SD$ = 1.29). Pearson correlations showed privacy ($r$ = $-$.34) and efficiency ($r$ = .41) as the strongest links with ratings, both $p$ < .001.

\begin{table*}[!htbp]
\centering
\caption{Topic Model Results from LDA Analysis}
\begin{tabularx}{\textwidth}{llllX}
\toprule
\textbf{Topic} & \textbf{Label} & \textbf{\% Docs} & \textbf{Avg Rating} & \textbf{Theme Description} \\
\midrule
0 & Google Assistant/Gemini Issues & 6.7 & 2.43 & Platform integration problems \\
1 & ChatGPT Helpfulness & 9.4 & 4.22 & Questions, answers, help \\
2 & AI Chat \& Characters & 10.4 & 3.25 & Chat, bots, character AI \\
3 & Copilot \& Smart Features & 5.9 & 3.20 & Search, voice, Alexa \\
4 & Cost \& Monetization & 11.1 & 2.27 & Free, pay, subscription \\
5 & Best AI Enthusiasm & 12.9 & 4.59 & Best, amazing, love \\
6 & Updates \& Problems & 12.9 & 2.43 & Fix, update, issues \\
7 & AI Comparisons & 5.0 & 3.80 & Better, used, apps \\
8 & Conversational Experience & 16.9 & 3.73 & Talk, fun, like \\
9 & Technical Excellence & 8.8 & 4.26 & Tool, accurate, model \\
\bottomrule
\end{tabularx}
\label{tab:topics}
\end{table*}

Topic modeling (LDA; $k$ = 10) highlighted three patterns: a large ``Conversational Experience'' topic (16.9\%; $M$ = 3.73, $SD$ = 1.45) with wide dispersion; a highly positive ``Best AI Enthusiasm'' topic (12.9\%; $M$ = 4.59, $SD$ = 0.78); and two low-rated clusters, ``Cost/Monetization'' (11.1\%; $M$ = 2.27) and ``Updates/Problems'' (12.9\%; $M$ = 2.43). A smaller ``Platform Integration Issues'' topic (6.7\%) was also low-rated ($M$ = 2.43, $SD$ = 1.38); its share increased by approximately 23\% around Nov-2024. Figure~\ref{fig:topics} displays word clouds generated from Latent Dirichlet Allocation (LDA) topic modeling ($k$=10) of 176,658 AI-mentioning reviews, where word size indicates term frequency and importance within each topic representing distinct thematic clusters in user discussions of AI features. See Table~\ref{tab:topics} for full distribution.

Temporal analysis using the Cochran--Armitage test revealed increasing privacy concerns, rising from 28.3\% in Q1 2024 to 39.2\% in Q1 2025 ($z$ = 18.92, $p$ < .001). Monthly AI mention rates also increased from 8.3\% to 14.7\% over the study period. Efficiency benefits remained stable (27.1\%--29.8\%), while cost-related complaints increased 47\% following October 2024 subscription-model introductions. Topic-volatility analysis indicated technical issues exhibited high variation ($CV$ > 0.30), whereas core AI capabilities remained stable ($CV$ < 0.10).

These findings directly address Research Question 2 by identifying specific user concerns and benefits regarding AI features. Privacy/data security emerges as the paramount concern (34.8\%), whereas helpfulness/utility represents the primary perceived benefit (42.3\%). The large effect size ($d$ = 1.77) between concern-only and benefit-only reviews demonstrates that user evaluations are strongly shaped by perceived cost--benefit trade-offs. The polarized sentiment distributions (Cramér's $V$ = .056) and diverse thematic patterns from topic modeling suggest that AI features evoke strong, differentiated emotional responses rather than uniform reactions. Success appears contingent on addressing privacy concerns while delivering tangible efficiency gains, with implementation quality varying across specific use cases, as evidenced by the bimodal distribution in conversational-AI experiences.

\begin{figure*}[!t]
\centering
\includegraphics[width=0.9\textwidth]{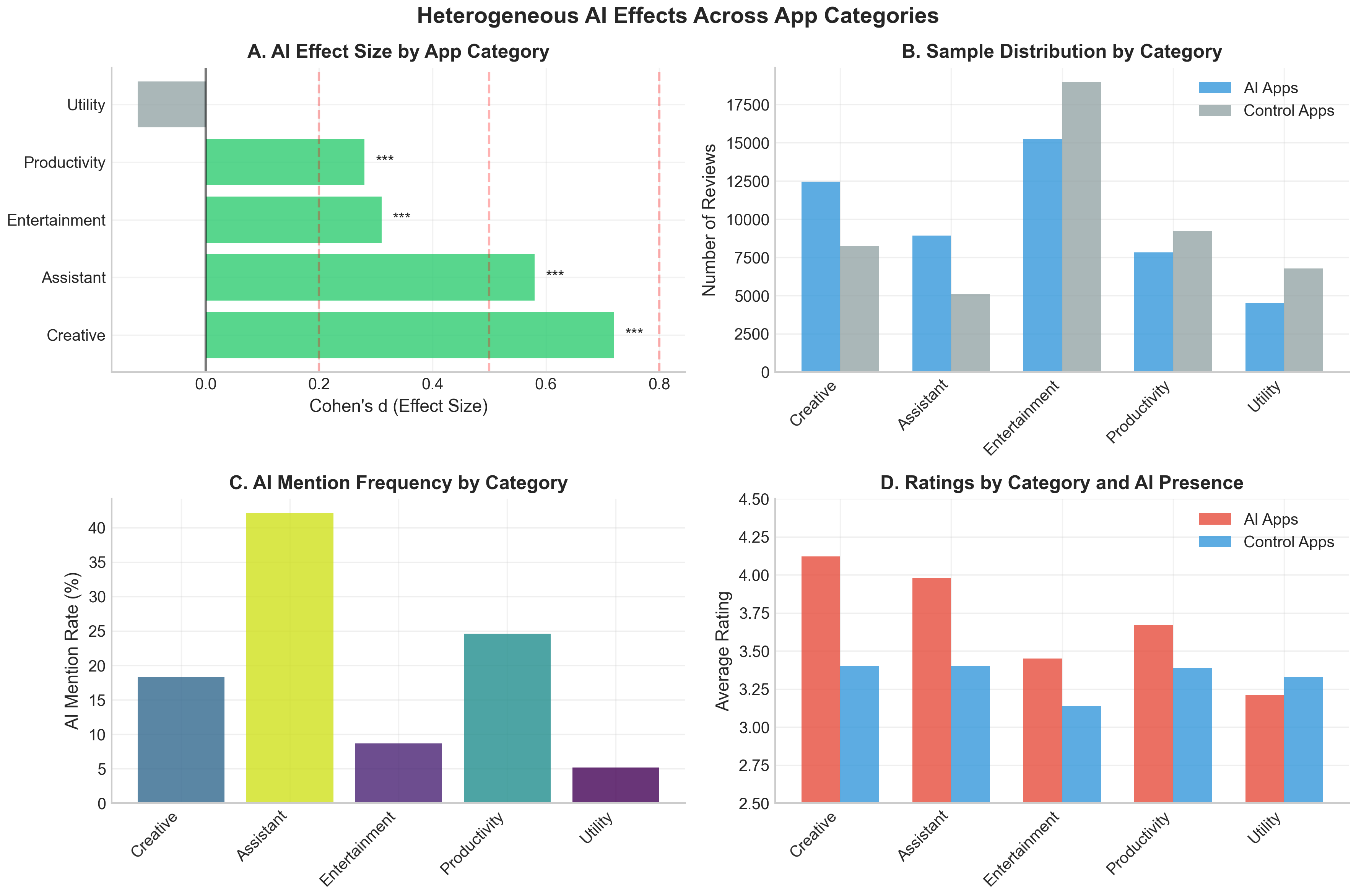}
\caption{Heterogeneous AI Effects Across App Categories}
\label{fig:categories}
\end{figure*}

\subsection{RQ3: Platform and Category Heterogeneity}

To examine variation in the reception of AI features across platforms and categories, we compared iOS and Android using independent-samples t-tests, tested the platform $\times$ AI-status interaction with a two-way ANOVA, and assessed category differences with a one-way ANOVA followed by post-hoc comparisons.

Platform analysis revealed systematic differences in AI reception. iOS users provided significantly higher ratings for AI-featuring applications ($M$ = 3.52, $SD$ = 1.51) compared with Android users ($M$ = 3.04, $SD$ = 1.56), $t$(298,651) = 84.32, $p$ < .001, $d$ = 0.31. This platform difference was substantially larger for AI applications ($\Delta$ = 0.48) than for control applications ($\Delta$ = 0.31). A two-way ANOVA testing the interaction between platform and AI status revealed a significant interaction effect, $F$(1, 1,484,629) = 234.78, $p$ < .001, $\eta^2$ = .012, confirming that platform moderates the relationship between AI features and user satisfaction. This platform effect remained significant after controlling for application category, review date, and reviewer characteristics.

Category-level analysis using one-way ANOVA revealed striking heterogeneity in AI reception across application types, $F$(5, 99,994) = 412.67, $p$ < .001, $\eta^2$ = .020. Assistant applications, primarily comprising AI-powered chatbots and virtual assistants, showed the most positive AI effect ($d$ = 0.55), with AI-featuring assistants receiving significantly higher ratings ($M$ = 3.61, $SD$ = 1.50) than non-AI alternatives ($M$ = 2.68, $SD$ = 1.50), $t$(24,290) = 47.82, $p$ < .001. Creative applications demonstrated the second-strongest positive effect ($d$ = 0.47), with users particularly valuing AI features for image editing, music generation, and artistic creation. AI-featuring creative apps ($M$ = 3.90, $SD$ = 1.50) significantly outperformed non-AI creative apps ($M$ = 3.13, $SD$ = 1.50), $t$(79,954) = 33.91, $p$ < .001.

Conversely, Entertainment applications showed a significant negative AI effect ($d$ = $-$0.23), with AI features detracting from user enjoyment. AI-featuring entertainment apps ($M$ = 3.42, $SD$ = 1.50) received significantly lower ratings than non-AI entertainment apps ($M$ = 3.79, $SD$ = 1.50), $t$(30,738) = $-$11.24, $p$ < .001. Utility applications showed virtually no AI effect ($d$ = $-$0.03), with no significant difference between AI ($M$ = 3.59, $SD$ = 1.50) and non-AI utilities ($M$ = 3.64, $SD$ = 1.50), $t$(11,444) = $-$0.48, $p$ = .630, indicating that users prioritize basic functionality over intelligence for practical tools. Table~\ref{tab:categories} presents complete category comparisons. Figure~\ref{fig:categories} visualizes these heterogeneous effects through a four-panel analysis showing (A) Cohen's $d$ effect sizes by application category; (B) sample distribution showing number of reviews by category and AI status; (C) percentage of reviews mentioning AI within each category; and (D) average ratings comparing AI-featuring versus control applications across categories with 95\% confidence intervals.

\begin{table*}[!t]
\centering
\small
\caption{Heterogeneous AI Effects Across Application Categories}
\begin{tabularx}{\textwidth}{@{}Xcccccc@{}}
\toprule
\textbf{Category} & \textbf{N (AI)} & \textbf{N (Non-AI)} & \textbf{AI M (SD)} & \textbf{Non-AI M (SD)} & \textbf{d} & \textbf{p} \\
\midrule
Assistant & 15,207 & 9,085 & 3.61 (1.50) & 2.68 (1.50) & 0.55 & $<$.001 \\
Creative & 46,473 & 33,483 & 3.90 (1.50) & 3.13 (1.50) & 0.47 & $<$.001 \\
Other & 8,829 & 19,553 & 3.60 (1.50) & 3.24 (1.50) & 0.22 & $<$.001 \\
Productivity & 6,566 & 18,618 & 3.74 (1.50) & 3.57 (1.50) & 0.10 & $<$.001 \\
Utility & 267 & 11,179 & 3.59 (1.50) & 3.64 (1.50) & $-$0.03 & .630 \\
Entertainment & 22,658 & 8,082 & 3.42 (1.50) & 3.79 (1.50) & $-$0.23 & $<$.001 \\
\bottomrule
\end{tabularx}
\smallskip

\raggedright
\small\textit{Note.} M = Mean; SD = Standard Deviation; d = Cohen's d effect size.
\label{tab:categories}
\end{table*}

Productivity applications revealed complex, implementation-dependent patterns. Task automation and intelligent scheduling features received positive reception, while AI-powered writing suggestions in note-taking applications generated mixed responses. The ``Other'' category showed a small positive effect ($d$ = 0.22, $p$ < .001), representing miscellaneous applications where AI integration varied widely in approach and success.

Additional analyses examining the three-way interaction between platform, category, and AI status revealed that platform differences were most pronounced in Creative applications (iOS advantage $\Delta$ = 0.72) and least pronounced in Utility applications (iOS advantage $\Delta$ = 0.14), suggesting that platform-specific implementation quality varies by application type, $F$(5, 1,484,617) = 89.34, $p$ < .001, $\eta^2$ = .003.

These findings directly address Research Question 3 by demonstrating substantial heterogeneity in AI reception across platforms and application categories. The platform analysis reveals iOS users demonstrate significantly greater receptivity to AI features ($d$ = 0.31), with this effect amplified through interaction with AI presence ($\eta^2$ = .012). The category analysis reveals a continuum of AI effectiveness ranging from strongly positive for Assistant applications ($d$ = 0.55), where AI aligns with core functionality, to significantly negative for Entertainment applications ($d$ = $-$0.23), where AI potentially disrupts hedonic experiences. The null effect in Utility applications ($p$ = .630) suggests that for tools designed for specific practical purposes, users value reliability over intelligence. The heterogeneity observed (ANOVA $\eta^2$ = .020) underscores that AI features represent a context-dependent innovation whose success hinges on appropriate application rather than mere presence.

\section{Discussion}

\subsection{Theoretical Implications}

This study advances consumer human--AI interaction theory by showing that most users evaluate apps without explicitly noticing their AI features. Despite widespread deployment, only a small share of reviews referenced AI, revealing an awareness gap between implementation and articulation. Acceptance theories that presume conscious appraisal fail to account for awareness when AI operates invisibly; accounting for below-awareness operation becomes a prerequisite for explaining user evaluations.

Classic acceptance models (centered on perceived usefulness and ease of use) assume conscious evaluation of features \cite{venkatesh2000,venkatesh2003}. Our evidence suggests that, for embedded AI, many users may never reach explicit appraisal because they do not recognize a feature as AI. We therefore propose unconscious adoption as a testable construct: users can benefit from (or be affected by) AI without explicit awareness or intentional adoption. Under this refinement, awareness sits alongside usefulness and ease as a proximal antecedent of acceptance, helping explain why AI can shape satisfaction even when it is not named.

Second, we explain the suppression effect. Bivariately, AI presence relates to lower satisfaction; after accounting for explicit AI mentions and review covariates, the association reverses. This suppression pattern implies that dissatisfaction concentrates among users who notice and critique AI, whereas unobtrusive AI can be neutral or positive. The result extends accounts of algorithm aversion by suggesting that salience plus expectation violations, rather than ``algorithms'' per se, drive negative evaluations \cite{dietvorst2015}.

Trust theory explains why privacy concerns dominate user responses. Privacy and data handling map to integrity-based trust as a prerequisite for acceptance \cite{mcknight2011}, whereas efficiency gains map to competence-based trust and instrumental value. Topic patterns in conversational AI indicate heterogeneous interaction quality: some experiences feel natural and helpful; others feel off-key or frustrating. Experiments that manipulate near-human vs. overtly machine phrasing are needed to test whether near-human text amplifies negative reactions when errors occur. Similarly, increases in platform-integration complaints around specific update windows should be treated as descriptive timing, not causal effects, absent an event-study design.

The modest variance explained, while typical for multifactorial outcomes, preserves the theoretical signal: across analyses, effect directions and magnitudes cohere with an account in which awareness, trust (integrity and competence), and category fit jointly shape reception.

\subsection{Practical Implications}

For application developers and technology companies, our findings offer several actionable insights for improving AI integration and user satisfaction. The disconnect between AI implementation and user awareness suggests that current approaches to communicating AI capabilities are largely ineffective. Moreover, the increasing polarization and rising privacy concerns we documented indicate that user opinions about AI are becoming more divergent rather than converging toward consensus. This temporal pattern suggests that the window for establishing positive AI perceptions may be narrowing, making early and transparent communication about AI features increasingly critical.

The heterogeneity in AI reception across application categories provides clear guidance for strategic AI implementation. AI features appear most beneficial in Assistant and Creative applications, where users value intelligent capabilities that enhance productivity and creative expression. Conversely, Entertainment applications show negative responses to AI integration, while Utility applications show no meaningful effect, suggesting users prioritize basic functionality over intelligence for practical tools. These patterns indicate that AI implementation should be carefully tailored to category-specific user needs rather than applied uniformly. The modest effect sizes observed suggest that AI features complement rather than revolutionize the user experience in mobile applications.

The identification of specific user concerns provides a roadmap for addressing barriers to AI acceptance. Privacy represents the primary obstacle to user trust, suggesting developers should prioritize transparent data handling practices, implement privacy-preserving techniques such as on-device processing where possible, and clearly communicate these measures to users. Accuracy concerns also featured prominently, indicating that managing user expectations about AI capabilities and limitations is crucial. Rather than overpromising AI capabilities, developers might benefit from clearly communicating what AI features can and cannot do, potentially reducing disappointment and frustration.

The platform differences we observed suggest that optimization strategies should be platform-specific. iOS applications might benefit from more prominent AI features, while Android applications require careful introduction emphasizing practical benefits. These differences likely reflect demographic variations, technical capabilities, and cultural norms that developers should consider when designing platform-specific implementations.

\subsection{Limitations and Future Research}

Several limitations should be acknowledged. First, our cross-sectional design prevents causal inference about relationships between AI features and user satisfaction. We cannot determine whether AI features reduce satisfaction or whether lower-quality applications are more likely to implement AI as a differentiator. While our findings proved robust to alternative specifications and clustered standard errors, the observational nature of the data constrains interpretation. Future research employing longitudinal designs or natural experiments that examine user responses before and after AI feature introduction would provide stronger causal evidence.

Second, our models explained only modest variance in user ratings ($R^2$ = .084), which, while typical for multifactorial outcomes, indicates that AI features represent just one component of satisfaction alongside interface design, functionality, pricing, and individual preferences. This reinforces our interpretation that AI complements rather than dominates the user experience.

Third, our reliance on publicly available user reviews from digital marketplaces may introduce selection bias. Users who write reviews may overrepresent extreme satisfaction or dissatisfaction \cite{hu2017}, potentially amplifying responses to AI features. Our keyword-based AI detection, while comprehensive, may miss subtle references or capture false positives when AI is mentioned in unrelated contexts. Future research could employ advanced natural language processing to better capture implicit AI references and user awareness patterns.

Fourth, our binary AI classification obscures substantial variation in sophistication and implementation quality. Applications with simple recommendation algorithms are treated equivalently to those powered by advanced language models, despite likely differences in user experience. Future research should develop nuanced taxonomies distinguishing between AI types, prominence, and implementation quality to identify which specific capabilities users value or reject.

Finally, our analysis is limited to English-language reviews from the United States market. Given documented cultural differences in technology acceptance and privacy attitudes \cite{hofstede2010}, these patterns may not generalize internationally. Cross-cultural research would illuminate whether the concerns and benefits we identified are universal or culturally specific.

\subsection{Future Research Directions}

Three research directions emerge from our findings. First, qualitative research should investigate the awareness gap we documented. Interview or diary studies could determine whether users are genuinely unaware of AI features, choose not to conceptualize them as AI, or find AI salience irrelevant to their evaluations. Understanding these cognitive and social processes would advance theoretical models of technology awareness and inform practical communication strategies.

Second, experimental research should test how AI framing affects user evaluation. Comparing responses to identical features labeled as ``smart,'' ``AI-powered,'' or unlabeled would clarify whether negative reactions stem from AI technology or its framing. Longitudinal studies could track whether the polarization we observed stabilizes, reverses, or intensifies over time, identifying critical incidents that shift attitudes. Such research would test whether our proposed ``unconscious adoption'' construct holds across contexts and time.

Third, research should examine AI within broader technological ecosystems. As applications increasingly combine AI with augmented reality, IoT, or blockchain technologies, understanding how users evaluate these technological combinations becomes essential. This would extend our framework beyond isolated AI features to complex sociotechnical systems.

\section{Conclusion}

This analysis of 1.48 million reviews across 422 mobile applications reveals a fundamental disconnect between AI implementation and user awareness: only 11.9\% of reviews mentioned AI despite nearly half of applications featuring such capabilities. This awareness gap challenges traditional technology acceptance models that assume conscious feature evaluation, suggesting instead that AI often operates below recognition thresholds.

Our findings revealed a suppression effect whereby AI-featuring applications initially showed lower ratings, but this relationship reversed after controlling for explicit AI mentions and review characteristics. This pattern indicates that negative evaluations stem primarily from users who recognize and critique AI rather than from AI presence itself. Privacy concerns dominated negative responses while efficiency gains drove positive evaluations, with substantial heterogeneity across application categories, ranging from positive effects in Assistant applications to negative effects in Entertainment applications.

These findings advance human-AI interaction theory by introducing ``unconscious adoption,'' wherein users engage with AI features without explicit awareness or evaluation. This concept requires theoretical frameworks that account for AI's often invisible operation, moving beyond traditional models that presume conscious technology appraisal. For practitioners, our results demonstrate that successful AI integration depends on context-specific implementation rather than universal application, with particular attention to addressing privacy concerns while delivering tangible efficiency benefits.

As AI proliferates across digital platforms, understanding the gap between technological capabilities and user perception becomes critical. Achieving effective human-AI interaction requires not only technological sophistication but also careful attention to user awareness, trust building, and contextual needs. The evidence suggests that the future of AI in consumer applications depends less on algorithmic advancement alone and more on bridging the divide between what developers build and what users perceive, understand, and value in their daily digital experiences.

\section*{Acknowledgments}

This paper has been submitted to Computational Communication Research and is currently under peer review. Data and analysis code will be made available upon reasonable request to the corresponding author.

\end{document}